# The effect of extended closure of red-light areas on COVID-19 transmission in India

Abhishek Pandey[1]*, Sudhakar V. Nuti[2], Pratha Sah[1], Chad Wells[1], Alison P. Galvani[1],

Jeffrey P. Townsend[3]

[1]Center for Infectious Disease Modeling & Analysis, Yale University, New Haven, CT.

[2]Department of Medicine, Massachusetts General Hospital and Harvard Medical School, Boston, MA.

[3]Department of Biostatistics, Yale School of Public Health, New Haven, CT.

*corresponding author

**Abstract**

The novel coronavirus disease (COVID-19) pandemic has resulted in over 200,000 cases in India. Thus far, India has implemented lockdown measures to curb disease transmission. However, commercial sex work in red-light areas (RLAs) has potential to lead to COVID-19 resurgence after lockdown. We developed a model of COVID-19 transmission in RLAs, evaluating the impact of extended RLA closure compared with RLA reopening on cases, hospitalizations, and mortality rates within the RLAs of five major Indian cities, within the cities, and across India. Closure lowered transmission at all scales. More than 90% of cumulative cases and deaths among RLA residents of Kolkata, Pune, and Nagpur could be averted by the time the epidemic would peak under a re-opening scenario. Across India, extended closure of RLAs would benefit the population at large, delaying the peak of COVID-19 cases by 8–23 days, and avert 32–60.2.% of cumulative cases and 43–67.6% of cumulative deaths at the peak of the epidemic. Extended closure of RLAs until better prevention and treatment strategies are developed would benefit public health in India.

**Introduction**

The novel coronavirus disease (COVID-19) pandemic, caused by the severe acute respiratory syndrome–coronavirus 2 (SARS-CoV-2) virus, has resulted in millions of cases, hundreds of thousands of deaths, and a negative economic impact worldwide. India, a country with over 1.3 billion people across metropolitan areas and rural villages and an underdeveloped medical infrastructure[1], could be particularly hard hit with the unmitigated spread of COVID-19[2]. To address this challenge, India has implemented widespread lockdown measures, including social distancing and travel restrictions[3,4]. On March 24, India first announced nationwide lockdown for three weeks, effectively home quarantining everyone in the country to curb the pandemic growth. The lockdown in India was subsequently extended three times, first to May 3, then to May 17, and currently to May 31[5]. Evidence from both India and abroad demonstrates that social distancing is essential to prevent the spread of COVID-19 and reduce mortality[1,6–10], especially until a vaccine is developed. Nevertheless, several countries, including India, are now cautiously beginning to lift some restrictions in hopes of restarting the economy and preventing economic distress.

The government of India has categorized districts of India into three zones based on the COVID-19 risks. The hotspots of transmission, categorized as "red zones," are identified based on total active cases, doubling rate of confirmed cases, extent of testing, and district feedback. Areas with declining or stable numbers of cases are classified as "orange zones" and areas with no reported cases for a significant number of days are classified as "green zones"[11,12]. While nationwide lockdown continues until May 31, considerable relaxations for economic and public service activities are now being allowed in lower-risk districts marked as green and orange zones. During the first phase of reopening of the country—starting June 1—intra-state and inter-state travel will gradually be allowed without need of prior permission from the government. Similarly, places of worship, hotels, restaurants, malls and other hospitality services would resume operation from June 8[13]. As restrictions are eased within specific zones, attention should be directed to geographic hot spots that may disproportionately exacerbate the spread of COVID-19[14].

Red light areas (RLAs), where thousands of sex workers typically live and work[15], are one area of concern for rapid transmission of COVID-19. By design, these areas have high contact rates between sex workers and clients, and sex acts are not amenable to social distancing. Sex workers are vulnerable to high rates of infectious diseases[16,17], experiencing particularly high rates of asymptomatic transmission of infections—a notable component of COVID-19 epidemiology. Moreover, visitors to RLAs include many truck drivers and migrant workers[18], who not only live locally but travel long-distances and can potentially spread the virus more broadly, including to green and orange zones. The combined features of a high volume of visitors, high contact rates, potential higher infectivity of sex workers, and long-distance travel of clients across India may make the reopening of RLAs a risk to increasing COVID-19 transmission, health care utilization, and death. Therefore, the impact of COVID-19 within RLAs, on the cities in which they reside, and on the Indian populace requires critical evaluation. An analysis in Japan has demonstrated a surge of COVID-19 cases transmitted in RLAs—cases that have overwhelmed local hospitals[19]. Considering the high risk of COVID-19 transmission, other countries, such as the Netherlands[20], Germany[21], and Australia[22], have identified brothels as the last enterprises to reopen. In Australia, brothels and strip clubs are the only businesses to be designated as indefinitely closed.

Prior studies have evaluated the benefits of lockdown in India for slowing COVID-19 transmission[1,23,24]. However, no previous analysis has examined the effect that the reopening of RLAs would have on the spread of COVID-19 in India, or whether keeping them closed would lead to a reduction of cases, reduced health care utilization, and improved mortality rates. Such an analysis would be helpful to the national and local governments to make targeted decisions about when, where, and how to ease lockdown measures in the best interest of public health, the health care system, and the economy.

To understand the potential impact of extended closure of RLAs on COVID-19 in India, we developed a model that quantifies the effects of re-opening RLAs after the end of the lockdown. We estimated the change in the time to reach peak COVID-19 cases: the change in cases, hospitalization rates, and mortality rates; and the spread of COVID-19 within RLAs at both the national level and among some of the largest cities in India that have been designated within the red zones[11].

**Results**

*Study population*

Data collected on RLAs (**Table 1**) facilitated model parameterization. Closure of RLAs after lockdown significantly delayed the spread of COVID-19 in all cities and nationally, including reduced numbers of cases and deaths (**Fig. 2–3**). The magnitude of these effects varied with greater infectiousness (increasing $R_0$; **Appendix Tables 3–4**) and increased with a greater resident population of RLAs relative to the general population of the city and with a greater contact rate between the general population of the city and residents of the RLA (**Fig. 3; Table 1**).

*Delay of the peak of cases*

The initial nationwide lockdown is projected to substantially delay the peak of the epidemic for each city considered and India (**Fig. 2, Appendix Table 3).** Extended closure of RLA after the lockdown is lifted can further delay the epidemic peak further by at least 8 days and up to 23 days with an $R_0$ of 1.75–2.25 in India (**Appendix Table 3)**. There was variation between peak delays among cities. The smallest delay in the peak of cases with an extended closure of the RLA in Mumbai was a 9-day delay using $R_0 = 2.25$ (a 114-day delay to a 123-day delay; **Appendix Table 3**). The largest delay in the peak of cases with an extended closure of the RLA in Kolkata was 117 days—close to the delay that was produced by lockdown alone—using $R_0 = 1.75$ (a 126-day delay to a 243-day delay; **Appendix Table 3**).

*Reduction in cases and deaths*

We found that an extended closure of RLAs after the initial lockdown period would avert 32% to 60.2% of cumulative cases and 43% to 67.6% of cumulative deaths across India when compared at the date of the peak of epidemic under re-opening of RLAs (**Fig. 3, Appendix Table 4**). Among cities, these reductions of COVID-19 cases and deaths were at least 49.7% and 59.2% respectively for $R_0 = 2$ (**Fig. 3**). In Kolkata, Pune, and Nagpur, reductions in cumulative cases and deaths at the date of this peak were more than 90% for all $R_0$ considered (**Appendix Table 4**).

*Cases, hospitalization, and mortality within RLAs*

Extended closure of RLAs after the initial lockdown reduced cases, hospitalizations, and mortality within RLAs in accordance with potential $R_0$ values for COVID-19. With re-opening of the RLAs, 32.5% (207,408) to 44.9% (285,908) of all RLA residents were projected to be infected by COVID-19 by the peak of the epidemic in India (**Appendix Table 5**). By the same date under a scenario of extended closure of the RLA, the proportion of RLA residents infected would be between 12.9% to 30.5%. For $R_0 = 2$, the maximum reduction in cumulative cases at the peak of epidemic occurs within the RLA of Kolkata (from 8,436 cases to 42 cases; **Appendix Table 5**) and the minimum reduction occurs within RLA of Mumbai (from 2195 cases to 1104 cases; **Appendix Table 5**).

The cumulative number of hospitalizations and deaths in RLAs would be substantially reduced with extended closure of the RLAs. Across all RLAs in India, at least 39.2% of cumulative hospitalizations, 39.9% of cumulative ICU admissions, and 42.9% of cumulative deaths could be averted by the date of the peak of epidemic if the RLAs remain closed (**Fig. 4A–C, Appendix Table 6–8**). By staying closed after lockdown, RLAs located in Kolkata, Pune and Nagpur could avert all ICU admissions and at least 94.6% of their cumulative hospitalizations and at least 92.9% of cumulative deaths by the date of the epidemic peak under a scenario of RLA re-opening (**Fig. 4A–C, Appendix Table 6–8**). Impact of extended closure of the RLA in Mumbai resembled the national trend with a reduction of 55.9% in cumulative hospitalization, 50% in cumulative ICU admissions, and 58.8% reduction in cumulative deaths for $R_0 = 2$ (**Fig. 4A–C**).

*Burden on hospital capacity*

India has approximately 1.9 million hospital beds, 95 thousand ICU beds, and 48 thousand ventilators. Most of the beds and ventilators in India are concentrated in seven states—Uttar Pradesh (14.8%), Karnataka (13.8%), Maharashtra (12.2%), Tamil Nadu (8.1%), West Bengal (5.9%), Telangana (5.2%), and Kerala (5.2%)[25]. As a result of extended closure of RLAs after the initial lockdown, current hospital capacity would be reached on October 26 rather than October 15 2020 **(Fig. 5)**. Moreover, at the projected November 19 peak of cases, India would need 10 times more hospital capacity than current capacity, while under extended closure of RLAs, required hospital capacity would be 5.8 times higher (**Fig. 5**).

Indian central and state governments are adding additional beds on a daily basis to ramp-up the healthcare capacity. Under the scenario in which closure of RLAs is not extended, the high number of imminent cases and consequent demand for hospitalization/ICU admission and ventilator use rates will likely surpass India's peak medical resource capacity, especially in the vulnerable zones—leading to a higher mortality rate **(Fig. 5).**

**Discussion**

Our study demonstrates a beneficial impact of extended closure of RLAs in India compared with their re-opening on COVID-19 cases, hospitalization and mortality. Extended closure would delay the peak number of cases by 8–23 days and result in a 32.0–60.2% reduction in the cumulative number of COVID-19 cases nationally, when compared at the date of the epidemic peak under a scenario of re-opening the RLAs. There would also be a 43–67.6% reduction in the cumulative number of COVID-19-related deaths nationally. These benefits of extended closure of RLAs, including a delayed peak in cases, a reduced increase in cases, and a reduction in deaths were demonstrated in Mumbai, New Delhi, Pune, Nagpur, and Kolkata, as well as across India. Mumbai and Kolkata (at the two extremes of $R_0$ considered) produced the most disparate results across cities—a difference that can be attributed to the size of the resident populations of the RLAs relative to the general population of the city and to the contact rates between the general population of the city and residents of the RLA.

The lockdown, contact tracing, and other post-lockdown government interventions [26,27] can continue to suppress transmission and flatten the curve, but it is unlikely for the pandemic to be resolved until there is a vaccine for the population[28]. Vaccine development and widespread distribution throughout India may take at least 18 more months[29]. In the absence of efficacious treatments or vaccines for COVID-19, there are limited public health interventions that can substantially reduce COVID-19 cases and deaths when re-opening a country as large and diverse as India[30,31]. Extended closure of RLAs in India may be one of these interventions—and it is feasible. Given the disproportionate impact of RLAs on COVID-19 transmission and the increase of mortality associated with its spread, extension of closure is essential to the protection of sex workers; their clients; the people who interact closely with sex workers and those close to RLAs, including local businesses, police personnel, NGO workers, and the local community; and the population of India at large. In addition to the lower immediate cases, hospitalizations, and deaths, extended closure confers additional time for the nation to plan and execute measures to protect public health and the economy, and to exchange public health and medical advances with the rest of the world. Similar to decisions to close cinema halls, gyms, and large public gatherings, RLAs should be critically evaluated for their ongoing potential to accelerate COVID-19 transmission and spread.

The outcomes of our model are supported by the experiences of other countries with COVID-19 and RLAs. In Japan, for example, medical facilities were overwhelmed by a surge of cases linked to an RLA[19,32,33]. The sharp increase in cases manifested among sex workers and their clients, and was largely contained within that sector only because of targeted and robust public health interventions. Japanese medical institutes have placed sex workers in the highest risk category for contracting the virus—the only profession in that classification not related to the medical field[34]. In Germany and Australia, brothels remain indefinitely closed, with some politicians calling for their permanent closure in Germany[35]. Due to concern regarding COVID-19 transmission, sole-operator sex workers and strip clubs have also been banned in Australia[21,22]. The diversity of businesses that function to enable commercial sex work or other activities involving close physical proximity as part of the nature of service share many of the same risk factors as the sex workers. These other businesses include strip clubs, ladies' bars, hotels that also commerce in sexual services, private sex-work establishments, spas, and massage parlors.

There are many social, economic, and health challenges, alongside the spread of COVID-19, that sex workers and their families will face under extended closure. Residents of RLAs typically live in confined, communal living spaces. Without sex work, they have very limited access to food and other vital living supplies. Furthermore, many sex workers lack government-approved documentation and thus are unable to benefit from the government's financial relief packages[36,37]. Toward this end, government action to extend closure of RLAs should be matched with commensurate programs that 1) extend payments to sex workers as part of the government's financial relief scheme for low-income individuals during the COVID-19 crisis, 2) ensure sex workers have access to additional resources via robust financial institutions that visibly offer better alternatives to high-interest lending schemes that entrap them in debt bondage, and 3) provide opportunities for sex workers to gain skills that facilitate their empowerment and adoption of lower-risk occupations. Over the longer term, educational and reintegration expenditures could be offset by profits generated via the redevelopment of RLAs. The COVID-19 crisis highlights the oft-neglected plight of sex workers and the global health burden that the RLAs impose in the context of infectious diseases. These global health burdens are also associated with extensive socioeconomic tolls. With regard to averting the public health and socioeconomic burdens, it would be beneficial for RLAs to remain closed until COVID-19 prevention and treatment strategies are developed and distributed.

Our findings are directly applicable to the five cities examined and to India as a nation. The Indian states that have the largest number of sex workers—Andhra Pradesh, Karnataka, Telangana, West Bengal, Maharashtra, New Delhi, Tamil Nadu, Madhya Pradesh, Gujarat, Uttar Pradesh, Rajasthan, and Kerala—are also those most affected by the current COVID-19 outbreak, which makes these findings especially pertinent to those locations[38]. Although we have not quantified the effect of extended closure in other localities within India, there is little reason to believe that RLAs or concentrated sex work will have qualitatively different effects in other locales. Considering the potential risk of COVID-19 transmission and spread to sex workers, their clients, and the broader community, policymakers in other countries that have large numbers of sex workers in RLAs—such as Nepal, Bangladesh, Brazil, the Democratic Republic of Congo, Cambodia, and Thailand—could also benefit via an extended closure of RLAs to prevent extensive, distributed transmission of COVID-19 and to preserve public health.

**Conclusion**

Our study predicts a significant increase in the surge in COVID-19 cases, hospitalizations, and deaths in India if RLAs are not closed post-lockdown, which could overwhelm the medical system, economy, and country. Compared with re-opening RLAs, extended closure of RLAs could reduce COVID-19 cases by 32–60.2% percent in cumulative cases and reduce cumulative COVID-19 deaths by 44–67.6% at the peak of epidemic. Moreover, extension of closure of RLAs after lockdown could delay the peak of COVID-19 cases in India by 8–23 days. With inevitable improvements over time in the implementation of contact tracing and testing—and eventually, the development of effective treatments and vaccination—this gained time would save many lives. These saved lives are a powerful argument for keeping RLAs closed in India until better prevention and treatment strategies for COVID-19 are available.

**Methods**

We developed an age-structured dynamic ordinary differential equation model for COVID-19 transmission that quantifies the contribution of RLAs towards the COVID-19 burden in India. Specifically, we assessed the impact of extended closure of RLAs after an initial country-wide lockdown in five cities of India (Mumbai, Nagpur, Delhi, Kolkata, and Pune) and nationally. A primary lockdown period lasted 3 weeks from 24 March 2020 to 14 April 2020, with subsequent continuations to 1 May 2020, 17 May 2020, and then to 31 May 2020[5]. The population of each location considered was compartmentalized into RLA residents and the general population. RLA residents included sex workers and non-sex workers including brothel managers, security, and support staff. Both populations were stratified into four age groups: 0–19 y, 20–49 y, 50–64 y, and ≥65 y. Age-distribution of each location was based on the most recent census[39], adjusted to current population estimates for five major cities and the RLAs within them, as well as the country of India. The general population and RLAs were further compartmentalized (**Fig. 1; Appendix Table 1**) based on the known natural history of COVID-19 disease as well as hospitalization and provision of intensive care units (ICUs).

In our model, a susceptible individual ($S$) that acquired infection remained in an incubation period ($E$) for an average of $1/\sigma$ = 5.2 days (**Appendix Table 2**). Individuals become infectious $1/\tilde{\sigma}$ = 2.3 days prior to symptom onset[40], moving to a pre-symptomatic infectious compartment ($E_I$) for that duration. Following this incubation period, an infected individual either remained asymptomatic ($I_A$) or developed symptoms ($I_N, I_H$). A proportion of symptomatic individuals ($(1-h)$) developed only mild symptoms ($I_N$). Symptomatic individuals with mild symptoms ($I_N, Q_N$) did not need hospitalization, and recovered in an average of $1/\gamma$ = 4.6 days **(Appendix Table 2)**. A proportion of individuals ($q$ = 0.05) with mild or severe symptoms were isolated ($E_I \rightarrow Q_N$, $E_I \rightarrow Q_H$). Symptomatic individuals with severe symptoms ($I_H, Q_H$) were either hospitalized ($H$), or required an ICU admission within a hospital ($C$). Those hospitalized patients ($H, C$) either recovered or died.

The spread of infection within each population depended on the prevalence of infection at the given time, age-specific contact patterns, and the per-contact transmission rate of the virus. In this model, the force of infection

$$\lambda = \frac{BC_M[M_A(E_I + k_A I_A + k_M I_N + I_H) + M_H(k_M Q_N + Q_H)]}{P}, \quad (1)$$

where $B$ is a matrix representing the probability of infection given contact within the general population and the RLA as well as the probability of infection given contact between the two subpopulations. Probability of infection within the RLA or within in the general population, $\beta$, was calibrated to the basic reproduction number (**Appendix Table 2**). The probability of infection given a contact between a client from the general population and a resident of an RLA was assumed to be one, if the resident was a sex-worker, and $\beta$, if the resident was a non-sex worker. As non-sex workers account for five times more interactions with clients than sex-workers, we calculated the probability of infection given a contact between a client from the general population and any resident of the red-light area as the weighted average of these two probabilities,

$$\beta_R = \frac{5\beta + 1}{6}. \tag{2}$$

Thus,

$$B = \begin{bmatrix} \beta & \beta_R \\ \beta_R & \beta \end{bmatrix}. \tag{3}$$

The interactions between the general population and the RLA occurring via clients were defined by the connectivity matrix

$$C_M = \begin{bmatrix} 1 & c_r \\ c_r & 1 \end{bmatrix}, \tag{4}$$

where $c_r$ is the contact rate between the two communities. This contact rate was calculated as the per-capita daily clients from the general population who visit the red-light area.

We used social contact matrices estimated for India overall and within specific locations such as households[41] to construct contact patterns between age-groups based on whether individuals are quarantined in their home or not. The contact patterns between age-groups were captured by two matrices:

$$M_A = \begin{array}{c|cccc} & \mathbf{0\text{--}19} & \mathbf{20\text{--}49} & \mathbf{50\text{--}64} & \mathbf{\geq 65} \\ \mathbf{0\text{--}19} & 14.4405 & 4.9482 & 1.6084 & 0.7058 \\ \mathbf{20\text{--}49} & 4.9482 & 9.2541 & 2.6860 & 0.6868 \\ \mathbf{50\text{--}64} & 1.6084 & 2.6860 & 1.4837 & 0.3820 \\ \mathbf{\geq 65} & 0.7058 & 0.6868 & 0.3820 & 0.3252 \end{array} \tag{5}$$

when individuals are not isolated / quarantined in their home ($I_N, I_H$), and

$$M_H = \begin{array}{c|cccc} & \mathbf{0\text{--}19} & \mathbf{20\text{--}49} & \mathbf{50\text{--}64} & \mathbf{\geq 65} \\ \mathbf{0\text{--}19} & 2.0309 & 2.0538 & 0.7890 & 0.5312 \\ \mathbf{20\text{--}49} & 2.0538 & 1.3993 & 0.7640 & 0.4330 \\ \mathbf{50\text{--}64} & 0.7890 & 0.7640 & 0.3975 & 0.2051 \\ \mathbf{\geq 65} & 0.5312 & 0.4330 & 0.2051 & 0.1434 \end{array} \qquad (6)$$

when they are ($Q_N, Q_H$) to match contact patterns at the household level[41].

Red-light areas in India are densely populated and have a high number of interactions when they are open. Therefore, for interactions between the general population and red-light area residents, we scaled up all entries in the contact matrix $M_A$ by a scaling factor $\rho$, calculated as the ratio of the average number of interactions that a client has per visit to the RLA (**Table 1)** to the average number of interactions in the contact matrix $M_A$.

We specified that individuals with asymptomatic and mild infections are only 50% infectious compared to severe infections (**Appendix Table 2**).

*Model fitting*

To generate epidemic projections, we first estimated the initial prevalence of COVID-19 at the beginning of lockdown by calibrating our model to the cumulative number of symptomatic cases during the week of April 22 to 28, 2020[42,43] and a range of plausible reproduction numbers for India (**Appendix Table 2)** using the least-squares method. Using our calibrated model, we generated results under scenarios of no initial lockdown, initial lockdown followed by return to status quo, and initial lockdown followed by extended closure of the RLA.

*Implementation of initial lockdown*

To implement the 68-day national lockdown in our model, we specified that everyone remained at home, and their contact patterns were informed by the household matrix $M_H$ for the duration of lockdown. Moreover, we set the interaction rate $c_r$ between the general population and the RLA at zero during this period.

*Post-lockdown*

After the initial lockdown period, contact patterns were informed by the overall contact matrix $M_A$, and it was assumed that as a result of improved contact-tracing capacity achieved during lockdown, 50% of symptomatic cases were isolated after the lockdown period[10]. For the scenario of extended closure of the RLA after lockdown, we maintained the contact rate $c_r = 0$; with no extended closure, it returned to its original value.

*Range of $R_0$ Values*

The basic reproduction number, $R_0$, is the expected number of cases directly generated by an infected person in a completely susceptible population early in an epidemic, without public health intervention. For example, if early in an outbreak, a single individual typically develops the infection and passes it to 2 people, the $R_0$ is 2. If $R_0 > 1$, there will be an exponential spread of the infection. If $R_0 < 1$, the rate of infection spread will be lower and eventually stop.

Epidemics grow faster with higher $R_0$. In this study, we show a range of results based on $R_0$ values of 1.75 to 2.25 calculated in recent research on the COVID-19 pandemic [10,44,45].

*Data collection*

To obtain current estimates of city-level population data, we applied population growth rate reported by the United Nations Department of Economic and Social Affairs[46] to the population size estimates for each city from latest census data (2011) [39]. For RLAs, extensive review and evaluation of research articles (including published reports, books, journals, research papers, program agendas/assessments/summaries), press releases, and credible media reports[38,47–50] was conducted for ascertaining accurate estimates for number of sex workers, number of brothels, and the number of sex workers per brothel.

We acquired primary data to complement estimates from the literature and acquire information on the number of non-sex workers per brothel, number of clients per sex worker per day, number of sex workers crossed on the way to a brothel in the RLA per visit, number of brothel workers met inside brothels on an average red-light visit by a client per visit, total interaction with sex workers and staff by a client per visit, and sex-worker interaction with other sex workers per day. Respondents were identified and selected based on their work experience in RLAs and continued access/exposure to the primary sampling units that comprised of active/former sex workers, brothel managers, security, support staff and communities inhabiting in and around RLAs. The work experience/access/exposure to RLAs for the respondents ranged from a minimum of one year to 15 years. The respondents were identified based on their close engagement with RLA residents, local police, the city's municipal corporation, NGOs specifically working in the particular RLA, NGOs addressing broad issues relating to RLA, counselors, health service providers, worker's association active in RLA, local business, and shop owners.

Cumulatively, among all 5 RLAs, 552 individuals were approached: 147 sex workers, 87 brothel managers, 143 clients, 33 social workers/researchers, 103 community members, and 39 local business owners. Among these individuals, 180 completed follow-up in-depth face to face interviews at 5 RLAs conducted in local languages resulting in a 32.6% overall response rate. The primary sample units include 48 sex workers, 31 brothel managers, 43 clients, 14 social workers/researchers, 24 community members, 20 local business owners. Trained field data collectors conducted confidential in-depth interviews with the sex workers after obtaining consent to share information. To estimate the population working in RLAs, non-sex-workers in RLAs were oriented to work as survey enumerators. The survey included demographic details, indicators of mobility, socio-economic vulnerability, engagement with clients, and routine activity patterns. The respondents' identities were kept confidential for safety reasons.

For national-level data, the number of sex workers, brothels, and client visits was determined from secondary sources[51,52]. Exhaustive face validation with subject experts was conducted for the dynamic data sets pertaining to the movement of sex / non-sex workers, clients, and their interaction within the brothels due to the high volatility of movement patterns of primary respondents at any given time-space in RLAs. Where more general secondary sources exhibited discrepancies with the specific RLA surveys, the more specific estimates from the five RLA surveys were used to compose final data at the national level.

**Data Availability**

References for all the data used in the analysis are provided in the article and supplementary material. All data generated from this study is shared publicly at the `Github` repository https://github.com/abhiganit/RedlLightAreas-COVID19

**Code Availability**

The mathematical model used to generate results for this study were developed and implemented in MATLAB. All code used for this study are publicly available at the `Github` repository https://github.com/abhiganit/RedlLightAreas-COVID19

# Figures

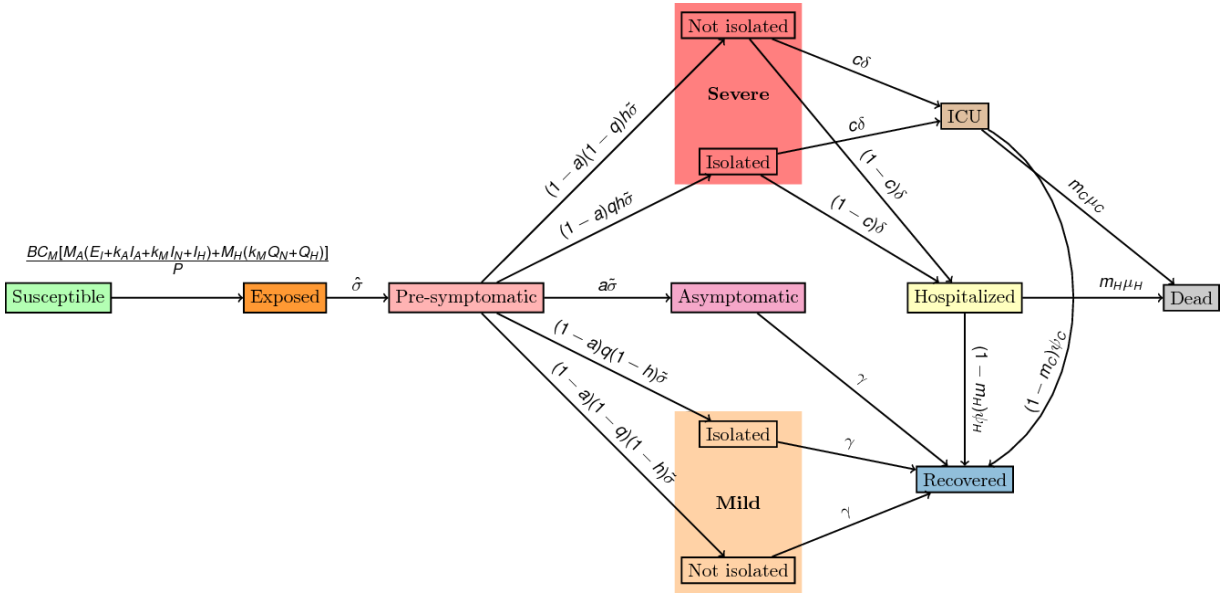

**Figure 1**. Model diagram. Schematic diagram of the COVID-19 transmission model.

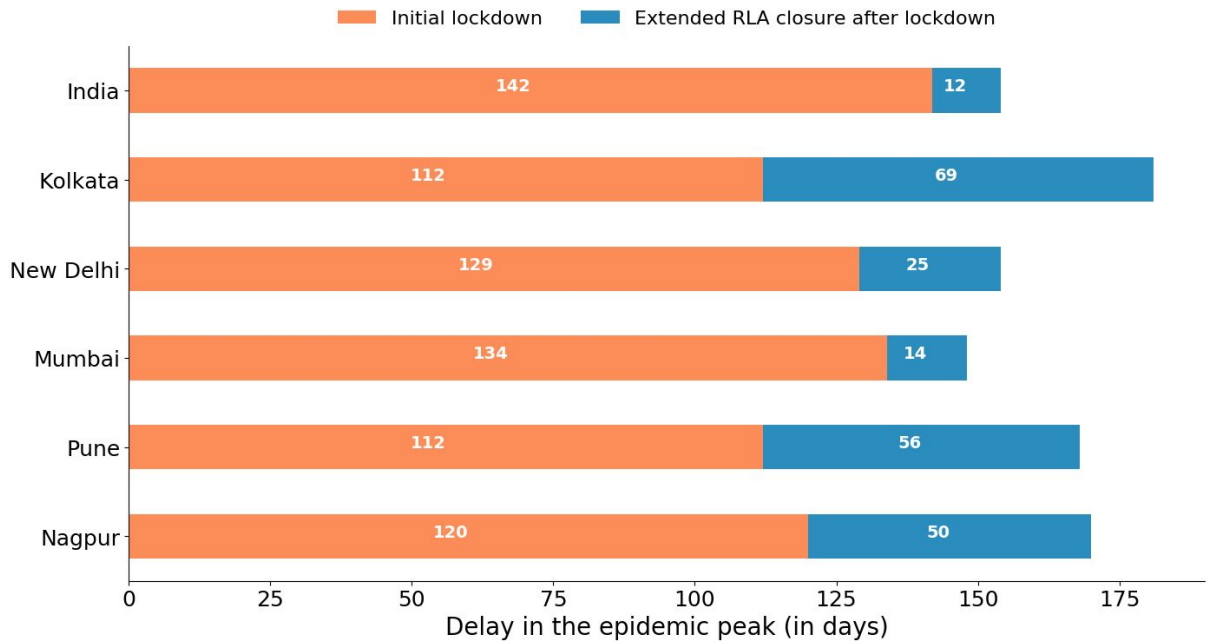

**Figure 2:** Delay in the epidemic peak. Delay (in days) in the peak of outbreak for each location as a result of initial lockdown from March 24 to May 31 and further delay if RLAs closure is extended at $R_0$ = 2.

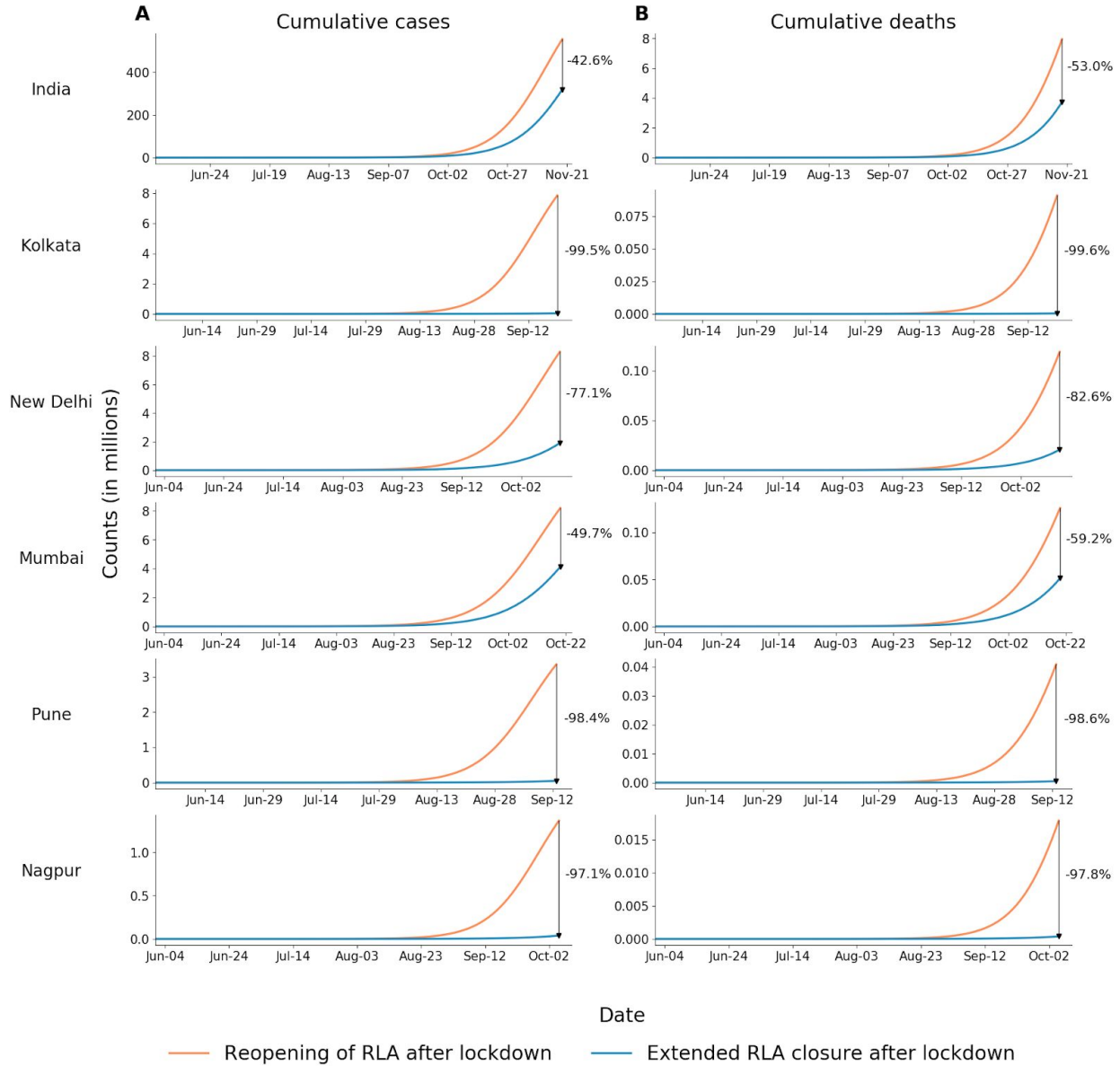

**Figure 3:** Cumulative cases and deaths at the epidemic peak. The effect of re-opening red-light areas vs. keeping them closed on COVID-19 cases and deaths in India, Kolkata, New Delhi, Mumbai, Pune and Nagpur, plotted from the date of the end of the initial lockdown until the date of the epidemic peak in that locality under the scenario of the re-opening of the red-light area (black arrow). **A**) Cumulative number of cases if the RLA remains closed (blue line) vs. the cumulative number of cases if the RLA re-opens after the lockdown (orange line). **B**) Cumulative number of deaths if the RLA remains closed (blue line) vs. the number of deaths if the RLA re-opens after the lockdown (orange line) at $R_0 = 2$.

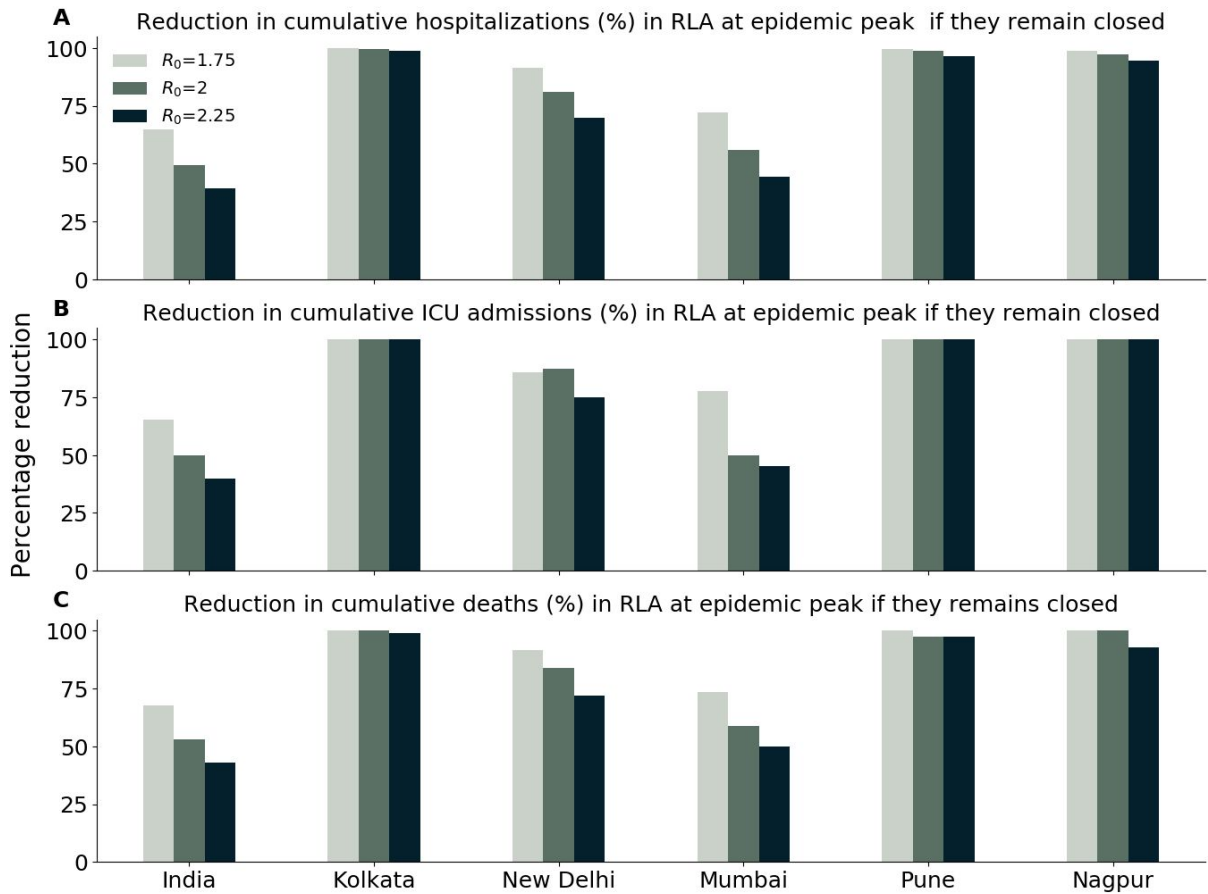

**Figure 4:** Reduction in COVID-19 burden in RLA as a result of their extended closure. Percentage reduction in **A**) cumulative hospitalizations, **B**) cumulative ICU admissions, and **C**) cumulative deaths as a result of extended closure of RLAs at the time the epidemic would peak if RLAs re-opened for each $R_0$ and location.

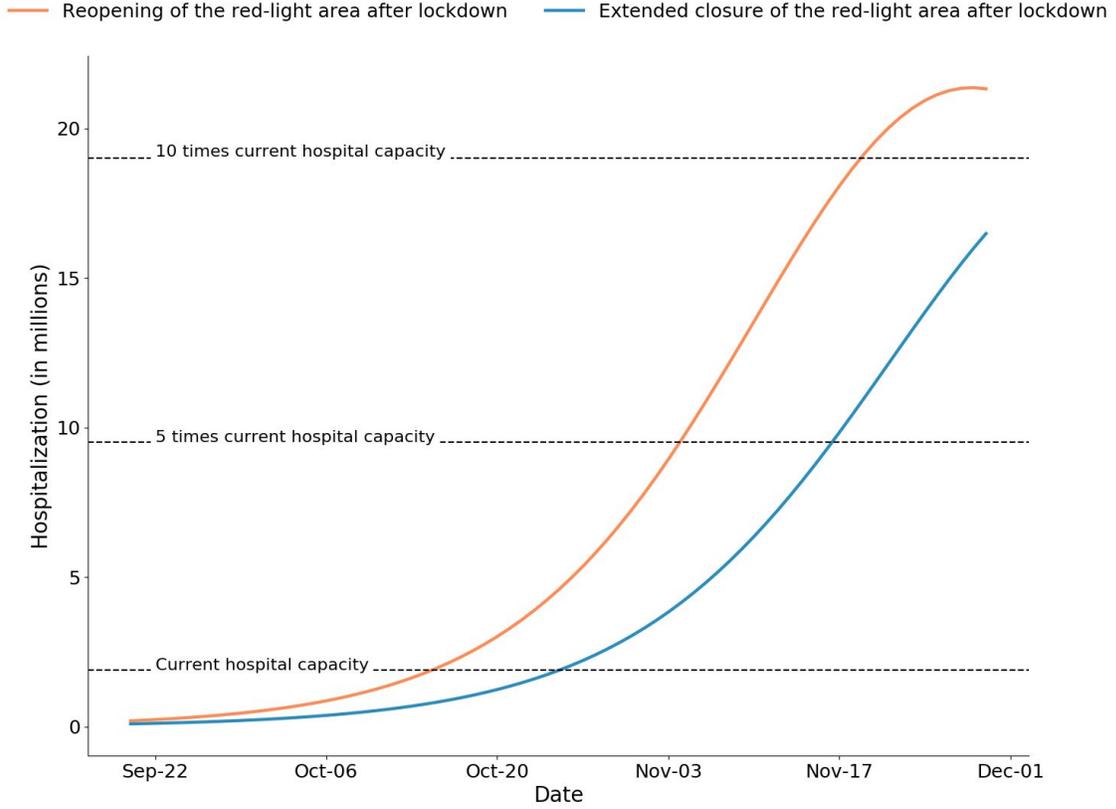

**Figure 5:** COVID-19 hospitalization over time in India. Number of hospitalizations over time after re-opening (orange) and extended closure (blue) of the RLAs in India, specifying $R_0$ = 2. Hospital capacity (dashed lines) is specified based on published estimates[25].

**Tables**

**Table 1**. Demography and red-light area data.

|  | India | Kolkata | New Delhi | Mumbai | Pune | Nagpur |
|---|---|---|---|---|---|---|
| General population (in thousands) | 1,380,004 | 14,850 | 19,500 | 20,411 | 6,629 | 2,893 |
| Red-light area population | 637,500 | 16,000 | 4,048 | 5,471 | 6,345 | 2,310 |
| Total daily interaction between clients and sex-workers in RLA | 585,000 | 33,000 | 10,500 | 9,000 | 10,000 | 4,200 |
| Contact rate between general population and RLA ($c_r$) | 0.00042 | 0.00222 | 0.00054 | 0.00044 | 0.00151 | 0.00145 |
| Total interactions with sex workers and staff by a client per visit | 35 | 64 | 74 | 49 | 82 | 60 |

29. Thompson, S. A. How long will a vaccine really take? *The New York Times* (2020).

30. Colbourn, T. COVID-19: extending or relaxing distancing control measures. *The Lancet. Public health* vol. 5 e236–e237 (2020).

31. Chatterjee, P. *et al.* The 2019 novel coronavirus disease (COVID-19) pandemic: A review of the current evidence. *Indian J. Med. Res.* **151**, 147–159 (2020).

32. Many people infected with Kabuki-cho customs such as cabaret clubs and hosts. *ANN News* https://news.tv-asahi.co.jp/news_society/articles/000180624.html (2020).

33. The Yomiuri Shimbun. Dozen-plus infected in Kabukicho, Tokyo's leading entertainment district. *The Japan News* https://the-japan-news.com/news/article/0006461421.

34. Surprising list of '23 occupations' with high corona risk. *Toyo Keizai* https://toyokeizai.net/articles/-/346397 (2020).

35. Deutsche Welle (www. dw.com). German lawmakers call for buying sex to be made permanently illegal | DW | 20.05.2020. *DW.COM* https://www.dw.com/en/german-lawmakers-call-for-buying-sex-to-be-made-permanently-illegal/a-53504221.

36. Ramaprasad, H. 'They are starving': women in India's sex industry struggle for survival. *The Guardian* (2020).

37. Tatke, S. India's sex workers fight for survival amid coronavirus lockdown. *Al Jazeera* https://www.aljazeera.com/news/2020/04/india-sex-workers-fight-survival-coronavirus-lockdown-200412073813464.html.

38. National AIDS Control Organization-Annual Report. http://naco.gov.in/sites/default/files/NACO%20ANNUAL%20REPORT%202016-17.pdf (2016).

39. MoHA, I. Census of India Website: Office of the Registrar General & Census Commissioner, India. (2011).
30

3232

# Appendix: The effect of extended closure of red-light areas on COVID-19 transmission in India

Abhishek Pandey[1]*, Sudhakar Nuti[2], Pratha Sah[1], Chad Wells[1], Alison P. Galvani[1],

Jeffrey P. Townsend[3]

[1]Center for Infectious Disease Modeling & Analysis, Yale University, New Haven, CT.

[2]Department of Medicine, Massachusetts General Hospital and Harvard Medical School, Boston, MA.

[3]Department of Biostatistics, Yale School of Public Health, New Haven, CT.

*corresponding author

**Appendix Table 1**. Compartments of the COVID-19 transmission model

| Compartment | Definition |
| --- | --- |
| $S$ | Susceptible |
| $E$ | Incubation |
| $E_I$ | Presymptomatic infections |
| $I_A$ | Asymptomatic infections |
| $I_H$ | Symptomatic severe infections (not isolated) |
| $I_N$ | Symptomatic mild infections (not isolated) |
| $Q_H$ | Symptomatic severe infections (isolated) |
| $Q_N$ | Symptomatic mild infections (isolated) |
| $H$ | Hospitalization |
| $C$ | Intensive care units |
| $D$ | Deaths |

**Appendix Table 2.** Model parameters

| Parameter | Definition | Value | Reference |
|---|---|---|---|
| $R_0$ | Reproduction number | 1.75–2.25 | 1,2 |
| $\beta$ | Probability of infection | Calibrated to $R_0$ | |
| $k_A$ | Relative infectivity of asymptomatic infections | 0.5 | 3 |
| $k_M$ | Relative infectivity of mild cases | 0.5 | 3 |
| $1/\sigma$ | Duration of incubation period | 5.2 | 4 |
| $1/\tilde{\sigma}$ | Duration of pre-symptomatic infectious period | 2.3 | 5 |
| $1/\hat{\sigma}$ | Duration of pre-symptomatic non-infectious period | $1/\sigma - 1/\tilde{\sigma}$ | |
| $a$ | Proportion of asymptomatic cases | 0.28 | 6 |
| $h$ | Proportion of severe symptomatic cases, age group 0–19 | 0.025 | 7 |
| | age group 20–49 | 0.32 | |
| | age group 50–64 | 0.32 | |
| | age group ≥65 | 0.64 | |
| $q$ | Proportion of symptomatic cases being isolated | 0.05 | 7 |
| $1/\gamma$ | Recovery period of mild and asymptomatic cases | 4.6 | 7 |
| $\delta$ | Hospitalization rate | 1/3.5 | 8 |
| $c$ | Proportion of symptomatic cases needing ICU in hospitals, age group 0–19 | 0.014 | 7 |
| | age group 20–49 | 0.042 | |
| | age group 50–64 | 0.075 | |
| | age group ≥65 | 0.15 | |
| $p_m$ | Proportion of hospitalized cases that die | 0.235 | 7 |
| $m_h$ | Coefficient matching model death rate among hospitalized (non-ICU) patients to observed probability of mortality. | $\dfrac{p_m \psi_H}{p_m \psi_H + (1 - p_m)\mu_H}$ | |
| $m_c$ | Coefficient matching the model death rate among ICU patients to the observed probability of mortality. | $\dfrac{p_m \psi_C}{p_m \psi_C + (1 - p_m)\mu_C}$ | |
| $1/\psi_H$ | Recovery period of hospitalized cases | 10 days | 9 |
| $1/\psi_C$ | Recovery period of hospitalized cases needing ICU | 13.25 days | 8 |
| $\mu_H$ | Mortality rate of hospitalized cases | 1/9.7 | 8 |
| $\mu_C$ | Mortality rate of hospitalized cases needing ICU | 1/7 | 10 |

**Appendix Table 3:** Delay (in days) in the peak of outbreak for each location and $R_0$.

|  | $R_0 = 1.75$ | | $R_0 = 2$ | | $R_0 = 2.25$ | |
| --- | --- | --- | --- | --- | --- | --- |
|  | L[a] | L + C[b] | L | L + C | L | L + C |
| India | 181 | 204 | 142 | 154 | 119 | 127 |
| Kolkata | 126 | 243 | 112 | 181 | 101 | 147 |
| New Delhi | 158 | 205 | 129 | 154 | 112 | 128 |
| Mumbai | 168 | 196 | 134 | 148 | 114 | 123 |
| Pune | 127 | 224 | 112 | 168 | 101 | 138 |
| Nagpur | 140 | 228 | 120 | 170 | 106 | 138 |

[a] Initial lockdown in India from 24 March 2020 to 31 May 2020 and reopening of red-light areas
[b] Extended closure of red-light areas after the initial lockdown.

**Appendix Table 4:** Percentage reduction in cumulative cases and deaths as a result of extended closure of RLAs at the time the epidemic would peak if RLAs reopened for each location and $R_0$.

|  | Percent reduction in cumulative cases | | | Percent reduction in cumulative deaths | | |
|---|---|---|---|---|---|---|
|  | $R_0 = 1.75$ | $R_0 = 2$ | $R_0 = 2.25$ | $R_0 = 1.75$ | $R_0 = 2$ | $R_0 = 2.25$ |
| India | 60.2% | 42.6% | 32% | 67.6% | 53% | 43% |
| Kolkata | 99.9% | 99.5% | 98.4% | 99.9% | 99.6% | 98.8% |
| New Delhi | 89.9% | 77.1% | 63.3% | 91.9% | 82.6% | 72.6% |
| Mumbai | 68.2% | 49.7% | 37.2% | 73.9% | 59.2% | 48.3% |
| Pune | 99.6% | 98.4% | 95.7% | 99.5% | 98.6% | 96.7% |
| Nagpur | 99.3% | 97.1% | 92.6% | 99.4% | 97.8% | 94.8% |

**Appendix Table 5:** Cumulative cases in the red-light area projected at the time the epidemic would peak if RLAs reopened.

| | RLA population | Cumulative cases | | | | | |
| --- | --- | --- | --- | --- | --- | --- | --- |
| | | $R_0 = 1.75$ | | $R_0 = 2$ | | $R_0 = 2.25$ | |
| | | L[a] | L + C[b] | L | L + C | L | L + C |
| India | 637,500 | 207408 | 82476 | 256099 | 147067 | 285908 | 194455 |
| Kolkata | 16,000 | 7739 | 8 | 8436 | 42 | 8990 | 144 |
| New Delhi | 4,048 | 1474 | 150 | 1724 | 396 | 1943 | 713 |
| Mumbai | 5,471 | 1756 | 558 | 2195 | 1104 | 2484 | 1559 |
| Pune | 6,345 | 2879 | 12 | 3207 | 51 | 3429 | 149 |
| Nagpur | 2,310 | 947 | 7 | 1089 | 32 | 1178 | 87 |

[a] Initial lockdown in India from 24 March 2020 to 31 May 2020 and reopening of red-light areas.
[b] Extended closure of red-light areas after the initial lockdown.

**Appendix Table 6:** Cumulative hospitalizations in the red-light area projected at the time the epidemic would peak under the scenario of RLA re-opening.

| | Cumulative hospitalization | | | | | |
|---|---|---|---|---|---|---|
| | $R_0 = 1.75$ | | $R_0 = 2$ | | $R_0 = 2.25$ | |
| | L[a] | L + C[b] | L | L + C | L | L + C |
| India | 20523 | 7199 | 24435 | 12402 | 26015 | 15817 |
| Kolkata | 752 | 1 | 795 | 3 | 822 | 10 |
| New Delhi | 155 | 13 | 174 | 33 | 189 | 57 |
| Mumbai | 187 | 52 | 227 | 100 | 247 | 137 |
| Pune | 285 | 1 | 310 | 4 | 320 | 11 |
| Nagpur | 96 | 1 | 107 | 3 | 112 | 6 |

[a] Initial lockdown in India from 24 March 2020 to 31 May 2020 and reopening of red-light areas.
[b] Extended closure of red-light areas after the initial lockdown.

**Appendix Table 7:** Cumulative ICU admissions in the red-light area projected at the time the epidemic would peak under the scenario of RLA re-opening.

| | Cumulative ICU admissions | | | | | |
|---|---|---|---|---|---|---|
| | $R_0 = 1.75$ | | $R_0 = 2$ | | $R_0 = 2.25$ | |
| | L[a] | L + C[b] | L | L + C | L | L + C |
| India | 924 | 320 | 1105 | 554 | 1179 | 709 |
| Kolkata | 34 | 0 | 36 | 0 | 38 | 0 |
| New Delhi | 7 | 1 | 8 | 1 | 8 | 2 |
| Mumbai | 9 | 2 | 10 | 5 | 11 | 6 |
| Pune | 13 | 0 | 14 | 0 | 15 | 0 |
| Nagpur | 4 | 0 | 5 | 0 | 5 | 0 |

[a] Initial lockdown in India from 24 March 2020 to 31 May 2020 and reopening of red-light areas.
[b] Extended closure of red-light areas after the initial lockdown.

**Appendix Table 8:** Cumulative deaths in the red-light area projected at the time the epidemic would peak under the scenario of RLA re-opening.

|  | Cumulative deaths | | | | | |
|---|---|---|---|---|---|---|
|  | $R_0 = 1.75$ | | $R_0 = 2$ | | $R_0 = 2.25$ | |
|  | L[a] | L + C[b] | L | L + C | L | L + C |
| India | 3402 | 1101 | 3667 | 1725 | 3542 | 2020 |
| Kolkata | 100 | 0 | 98 | 0 | 95 | 1 |
| New Delhi | 24 | 2 | 25 | 4 | 25 | 7 |
| Mumbai | 30 | 8 | 34 | 14 | 34 | 17 |
| Pune | 39 | 0 | 39 | 1 | 38 | 1 |
| Nagpur | 14 | 0 | 14 | 0 | 14 | 1 |

[a] Initial lockdown in India from 24 March 2020 to 31 May 2020 and reopening of red-light areas.
[b] Extended closure of red-light areas after the initial lockdown.